\documentclass[preprint,5p,times,twocolumn]{elsarticle}

\usepackage[
            CJKbookmarks=true,
            bookmarksnumbered=true,
            bookmarksopen=true,
            colorlinks=true,
            citecolor=cyan,
            linkcolor=red,
            anchorcolor=green,
            urlcolor=cyan
            ]{hyperref}

\usepackage{lipsum,amsmath,multicol}

\usepackage{subfigure}
\usepackage{mathrsfs}
\usepackage{multicol} 
\usepackage{graphicx,amsmath,amsfonts,amssymb,revsymb,dcolumn,epsfig,bm}
\usepackage{amsmath}
\usepackage{amssymb} 
\usepackage[utf8]{inputenc}

\begin{document}

\title{The longlived charged massive scalar field \\ in the higher-dimensional Reissner--Nordstr\"{o}m spacetime}
\author{Ming Zhang}\ead{mingzhang@mail.bnu.edu.cn}
\author{Jie Jiang}\ead{jiejiang@mail.bnu.edu.cn}
\author{Zhen Zhong \corref{cor1}} \ead{zhenzhong@mail.bnu.edu.cn}

\address{Department of Physics, Beijing Normal University, Beijing, 100875, China}

\cortext[cor1]{Corresponding author at: Department of Physics, Beijing Normal University,
Beijing, 100875, China}

\begin{abstract}
The quasinormal resonance frequency of the higher-dimensional Reissner--Nordstr\"{o}m (RN) black hole due to charged massive scalar field perturbation is deduced analytically in the eikonal regime. The characteristic decay timescale of the charged massive scalar perturbation in the background of the higher-dimensional RN spacetime is then obtained. The result reveals that longlived charged massive scalar field can exist in higher-dimensional RN spacetime under a certain condition.
\end{abstract}

 \maketitle

\section{Introduction}
The dynamics of the perturbation field in a spherically symmetric background spacetime is typically governed by a Schr\"{o}dinger-like equation together with proper boundary conditons at the horizon and the spatial infinity \cite{Berti:2009kk}. The quasinormal modes (QNMs), which is described as oscillations of the form $e^{-i\omega t}$, where $\omega$ is the quasinormal frequency, can be viewed as slowly decaying \cite{Nollert:1999ji,Cardoso:2001hn} or highly damped \cite{leaver1985ew,Nollert:1993zz,Hod:1998vk} waves travelling around the black hole. In fact, the quasinormal mode is nothing but the eigenvalue solution of the differential perturbation equation \cite{Kokkotas:1999bd}; in other words, the quasinormal mode can also be viewed as the Green’s function solution of the inhomogeneous wave equations \cite{Leaver:1986gd,Nollert:1992ifk,Berti:2006wq}. The investigations on the QNM has a 60-year long history, including the QNM's application on the recent discussions of Penrose strong cosmic censorship conjecture \cite{Cardoso:2017soq,Hod:2018dpx,Mo:2018nnu}.

The uniqueness theorem states that stationary black hole solutions originated from the Einstein's gravitational field equation can correspond to three conserved charges (mass, charge, and angular momentum) at most \cite{Carter:1971zc, Hawking:1971vc, Robinson:1975bv, Ipser:1971zz}. Motivated by the uniqueness theorem, it is stated that static matter field cannot be hold by the asymptotically flat black hole in Einstein's theory of gravitation by Wheeler's no-hair conjecture \cite{Ruffini:1971bza} and the dynamic purturbation around a static black hole will finally go to be vanished \cite{Mayo:1996mv,Nunez:1996xv,Bekenstein:1995un}; in particular, the scalar \cite{Bekenstein:1972ny}, massive vector \cite{Bekenstein:1971hc}, and spinor \cite{Hartle:1971qq} will be absorbed by the black hole or be scattered away from the black hole.

The Wheeler's no-hair conjecture, which holds except for hairy black hole \cite{Bizon:1990sr,Lavrelashvili:1992ia,Droz:1991cx}, however,  gives no implications of the decaying timescale that the black hole takes to absorb or scatter the purturbations. To this point, it has been analytically verified that the charged massive scalar field can be longlived (though vanished finally) outside the four-dimensional Reissner--Nordstr\"{o}m (RN)  black hole, as the relaxation time will be infinity under a certain condition $M\mu/qQ\rightarrow 1^-$ \cite{Hod:2016jqt}, where $M(\mu), Q(q)$ are respectively the mass and charge of the black hole (scalar field). Others researches about this topic can be seen in \cite{Konoplya:2018qov,Ferreira:2017cta, Toshmatov:2017qrq, Hod:2017gvn, Hod:2017lwb, Hod:2018pri, Hod:2017ehd, Hod:2017www, Hod:2016vkt, Hod:2016lgi}.

The researches of the perturbation and decay of fields around the higher-dimensional black hole have attracted tremendous interests \cite{Cardoso:2002pa,Konoplya:2003ii,Konoplya:2003dd,Berti:2003si,Cardoso:2003vt,Cardoso:2003qd}. In this paper, we will extend the investigation in \cite{Hod:2016jqt} to the case of the higher-dimensional RN black hole. In Sec. \ref{chaeq}, we will show the dynamic equation describing the scalar perturbation outside the higher-dimensional RN black hole. In Sec. \ref{qnmspe}, we will calculate the quasinormal resonance spectrum of the charged massive scalar perturbation. Lastly, the Sec. \ref{condis} will be devoted to our remarks.

\section{The equation of the charged massive scalar perturbation}\label{chaeq}
The metric of a $d$-dimensional RN black hole can be expressed as
\begin{equation}
ds^2=-f(r)dt^2+\frac{dr^2}{f(r)}+r^2 d\Omega^2_{d-2},
\end{equation}
where
\begin{equation}
f(r)=1-\frac{16 \pi  M r^{3-d}}{(d-2) \epsilon }+\frac{32 \pi ^2 Q^2 r^{-2 (d-3)}}{\left(d^2-5 d+6\right) \epsilon ^2},
\end{equation}
with $\epsilon= 2 \pi ^{\frac{\text{d}-1}{2}}/\Gamma \left((d-1)/2\right)$ the volume of the $(d-2)$-sphere, $M, Q$ are the mass and charge of the black hole, respectively. The electromagnetic field $F$ and gauge potential $A$ are
\begin{equation}
  F=(dA)_{ab},~A_a=\frac{4 \pi  Q r^{3-d}}{(3-d)\epsilon}dt.
\end{equation}
The Klein-Gordon wave equation describing the dynamics of a scalar field $\Psi (t,r,\Theta)$ with charge $q$ and mass $\mu$ can be written as
\begin{equation}
[(\triangledown^\nu-iqA^\nu)(\triangledown_\nu-iqA_\nu)-\mu^2]\Psi(t,r,\Theta)=0.
\end{equation}
Substituting the field decomposition 
\begin{equation}
  \Psi (t,r,\Theta)=\int\sum_{lm}e^{-i\omega t}\frac{R_{lm}(r, \omega)}{r^{\frac{d-2}{2}}}Y_{lm}(\theta)
\end{equation}
into the Klein-Gordon equation, one can obtain differential equations of $R_{lm}(r, \omega)$ and $Y_{lm}(\theta)$, which respectively reflect the radial and angular dynamics of the charged massive scalar field. After letting the two eigenfunctions share the same eigenvalue $K_l =l(d+l-3)$,where $l$ is the spherical harmonic index, we can write the radial function as
\begin{equation}\label{radial}
 f(r)^{2} R''(r)+f'(r)f(r) R'(r)+U R(r)=0,
\end{equation}
where
\begin{equation}
f'(r)\equiv \frac{\text{d}f(r)}{\text{d}r}, R'(r)\equiv \frac{\text{d}R(r)}{\text{d}r}, R''(r)\equiv \frac{\text{d}^2 R(r)}{\text{d}^2r}
\end{equation}
and
\begin{equation}\begin{aligned}
  U=&\left(\omega -\frac{4 \pi  q Q r^{3-d}}{(d-3) \epsilon }\right)^2-\frac{(d-4) (d-2) f(r)^2}{4 r^2}  \\&-\frac{f(r) \left[(d-2) r f'(r)+2 \left(K_l+\mu ^2 r^2\right)\right]}{2 r^2}.
\end{aligned}\end{equation}

As long as we define the tortoise coordinate $y$ by the differential relation
\begin{equation}
\text{d}y=\frac{\text{d}r}{f(r)},
\end{equation}
the radial equation (\ref{radial}) can be transformed into a Schr\"{o}dinger-like ordinary differential equation
\begin{equation}\label{sch}
\frac{\text{d}^2 R}{\text{d}y^2}+VR=0,
\end{equation}
where the effective radial potential $V$ can be written as
\begin{equation}\label{pote}
  V=\left(\omega -\frac{4 \pi  q Q r^{3-d}}{(d-3) \epsilon }\right)^2-f(r) H(r),
\end{equation}
with
\begin{equation}
H(r)=\mu^2+\frac{K_l}{r^2}+\frac{d^2-6 d+8}{4 r^2}+\frac{4 \pi  (d-2) M}{\epsilon  r^{d-1}}-\frac{8 \pi ^2 (3 d-8) Q^2}{(d-3) \epsilon ^2 r^{2 d-4}}.
\end{equation}

To solve (\ref{sch}) and obtain the quasinormal resonant frequency characterizing the relaxation dynamics of the charged massive scalar field, we should impose physically appropriate boundary conditions on the behaviours of the wave function. Specifically, there must be purely ingoing wave at the horizon of the black hole and purely outgoing wave at the spatial infinity \cite{Kodama:2003kk,Ishibashi:2003ap}, i.e.,
\begin{equation}\label{boundaryc}
R \backsim\left\{
\begin{aligned}
& e^{-i\left(\omega-\frac{4 \pi  q Q r^{3-d}}{(d-3) \epsilon }\right) y},~~~~~r\to r_+ ~(y\to -\infty); \\
& y^{-i \frac{4 \pi  q Q }{(d-3)\epsilon }} e^{i\sqrt{\omega^2 -\mu^2}  y},~~~~~r\to \infty ~(y\to \infty).
\end{aligned}
\right.
\end{equation}

With the boudary condition (\ref{boundaryc}), we can determine the quasinormal resonance spectrum $\left\{\omega _n(M,Q,\mu ,q,l)\right\}_{n=0}^{n=\infty }$ characterizing the dynamics of the charged massive scalar field around the $d$-dimensional RN black hole.

\section{The quasinormal resonance spectrum of the charged massive scalar perturbation}\label{qnmspe}
The horizon of a $d$-dimensional RN black hole is 
\begin{equation}\begin{aligned}
r_{+}=&(8 \pi )^{\frac{1}{d-3}} (d-3)^{\frac{1}{2 (3-d)}}\\&\times \left(\frac{\sqrt{2}\epsilon \sqrt{ \left(2 (d-3) M^2-(d-2) Q^2\right)}+2 \sqrt{d-3} M \epsilon }{Q^2}\right)^{\frac{1}{3-d}},
\end{aligned}\end{equation}
which implies that 
\begin{equation}
M^2 >\frac{d-2}{2(d-3)}Q^2,
\end{equation}
and
\begin{equation}
r_+ >(8 \pi )^{\frac{1}{d-3}} \left[\frac{M}{(d-2) \epsilon }\right]^{\frac{1}{d-3}}.
\end{equation}
Then, in the eikonal large-mass regime
\begin{equation}\label{eik}
  \mu\left(\frac{8\pi}{(d-2)\epsilon }M\right)^{\frac{1}{d-3}}\gg d+l-3,
\end{equation}
we can know that
\begin{equation}\begin{aligned}
H(r)&=\mu^2 (1+\frac{K_l}{\mu^2 r^2}+\frac{d^2-6 d+8}{4\mu^2 r^2})\\&~~~~+\mu^2(\frac{4 \pi  (d-2) M}{\mu^2 \epsilon  r^{d-1}}-\frac{8 \pi ^2 (3 d-8) Q^2}{\mu^2 (d-3) \epsilon ^2 r^{2 d-4}})\\&\sim \mu^2.
\end{aligned}\end{equation}
We have shown the specific reason of this approximation in \ref{appa}.

Then the radial potential (\ref{pote}) can be approximately written as
\begin{equation}\label{appveff}
V(r)=\left[\omega-\frac{4 \pi  q Q r^{3-d}}{(d-3) \epsilon } \right]^2-f(r)\mu^2\cdot \{1+\mathcal{O}[(\mu M^{\frac{1}{d-3}})^{-2}]\}.
\end{equation}
The radial potential can be viewed as an effective potential barrier, with its maximum value locating at  $r_0$ which can be obtained as
\begin{equation}
r_0=\left(\frac{(d-3) \epsilon \left[(d-2) q Q \omega -2 (d-3) \mu ^2 M\right]}{4 \pi  Q^2\left[(d-2) q^2-2 (d-3) \mu ^2\right]}\right)^{\frac{1}{3-d}}
\end{equation}
by the derivative of the coordinate $r$.

In the eikonal large-mass regime (\ref{eik}), using the method of WKB approximation, we can try to find the fundamental complex resonances determined by the Schr\"{o}dinger-like equation (\ref{sch}) and the effective radial potential (\ref{pote}) which is approximated to (\ref{appveff}). We can obtain the quasinormal frequency through the WKB resonance condition \cite{Konoplya:2003ii,Iyer:1986np,Schutz:1985zz}
\begin{equation}\label{wkb}
  \frac{V_0}{\sqrt{2V_0^{(2)}}}=-i\left(n+\frac{1}{2}\right)
\end{equation}
in the eikonal regime, where $V_0$ is the effective radial potential (\ref{appveff}) in the large-mass regime, and $V_0^{(k)}\equiv \text{d}^k V/\text{d}y^k$ ($k$ is a positive integer). Both $V_0$ and $V_0^{(k)}$ are evaluated at the extreme point $r_0$.

Specifically, the WKB resonance condition (\ref{wkb}) charaterizing the quasinormal frequency of the massve scalar field around the RN black hole can be expressed as
\begin{equation}\label{reso}
 \frac{\left[\omega-\frac{4 \pi  q Q r_0^{3-d}}{(d-3) \epsilon } \right]^2-\mu ^2 f\left(r_0\right)}{4 \sqrt{\pi } f\left(r_0\right) \sqrt{\frac{(d-3) r_0^{1-d} \left[(d-2) q Q \omega-2 (d-3) \mu ^2 M \right]}{(d-2) \epsilon }}}=-i(n+\frac{1}{2}),
\end{equation}
which cannot be directly solved analytically. However, we can decouple the real part $\omega_R$ and the imaginary part $\omega_I$ of the quasinormal frequency $\omega$ in the regime
\begin{equation}\label{rbiggeri}
\omega_R\gg \omega_I.
\end{equation}
For the real part of the WKB resonance condition, we can find the restriction equation as
\begin{equation}
\left[\omega_R -\frac{4 \pi  q Q r_0^{3-d}}{(d-3) \epsilon } \right]^2-\mu ^2 f\left(r_0\right)=0.
\end{equation}
Then the real part of the quasinormal frequency can be solved as
\begin{equation}
  \begin{aligned}
\omega_R=&\frac{M q}{Q}\\ &-\sqrt{\frac{2 (d-3) M^2-(d-2) Q^2}{2 \left(d^2-5 d+6\right) Q^2}}\times \sqrt{(d-2) q^2-2 (d-3) \mu ^2}.
\end{aligned}
\end{equation}
Meanwhile, from the resonance equation (\ref{reso}),  we can also get
\begin{equation}\begin{aligned}
  &2 \omega _I \left(\omega_R -\frac{4 \pi  q Q r_0^{3-d}}{(d-3) \epsilon }\right)=-4 \sqrt{\pi } \left(n+\frac{1}{2}\right) f(r_0)\\&\times\sqrt{\frac{(d-3) r_0^{1-d} \left[(d-2) q Q \omega_R -2 (d-3) \mu ^2 M \right]}{(d-2) \epsilon }},
\end{aligned}\end{equation}
from which it is not difficult to obtain the image part of the quasinormal resonance frequency as
\begin{equation}\begin{aligned}
\omega_I=&-\frac{2 \sqrt{\pi }f\left(r_0\right) \left(n+\frac{1}{2}\right)  }{\omega _R-\frac{4 \pi  q Q r_0^{3-d}}{(d-3)\epsilon }}\\&\times \sqrt{\frac{(d-3) r_0^{1-d} \left[(d-2) q Q \omega_R -2 (d-3) \mu ^2 M \right]}{(d-2) \epsilon }}.
\end{aligned}\end{equation}

Defining
\begin{equation}
\bar{Q}\equiv\frac{\sqrt{d-2}}{\sqrt{2(d-3)}}\frac{Q}{M},~\bar{\mu}\equiv\frac{\sqrt{2(d-3)}}{\sqrt{d-2}}\frac{\mu}{q},
\end{equation}
we can respectively express the real part and the image part of the quasinormal renonace frequency as
\begin{equation}\label{qnmr}
\omega_R =\frac{q-q \sqrt{\left(1-\bar{\mu }^2\right) \left(1-\bar{Q}^2\right)}}{\sqrt{2-\frac{2}{d-2}} \bar{Q}}
\end{equation}
and
\begin{equation}\label{qnmi}
\omega_I =\omega_I (\bar{\mu},\bar{Q}).
\end{equation}
We have shown concrete expressions of $\omega_I$ in \ref{appb}. It should be noticed that when $d=4$, our result is in consistent with that in \cite{Hod:2016jqt}.

\section{Remarks}\label{condis}
In this paper, we have analytically deduced the quasinormal resonance frequencies of the charged scalar field around the higher-dimensional RN black holes in the eikonal regimes (\ref{eik}). The quasinormal resonance mode can be written as
\begin{equation}\label{ome}
\omega=\frac{q-q \sqrt{\left(1-\bar{\mu }^2\right) \left(1-\bar{Q}^2\right)}}{\sqrt{2-\frac{2}{d-2}} \bar{Q}}-i\omega_I (\bar{\mu},\bar{Q}),
\end{equation}
where the concrete expressions of $\omega_I (\bar{\mu},\bar{Q})$ has been shown in \ref{appb}.

\begin{figure}[!htbp] 
   \centering
   \includegraphics[width=3.4in]{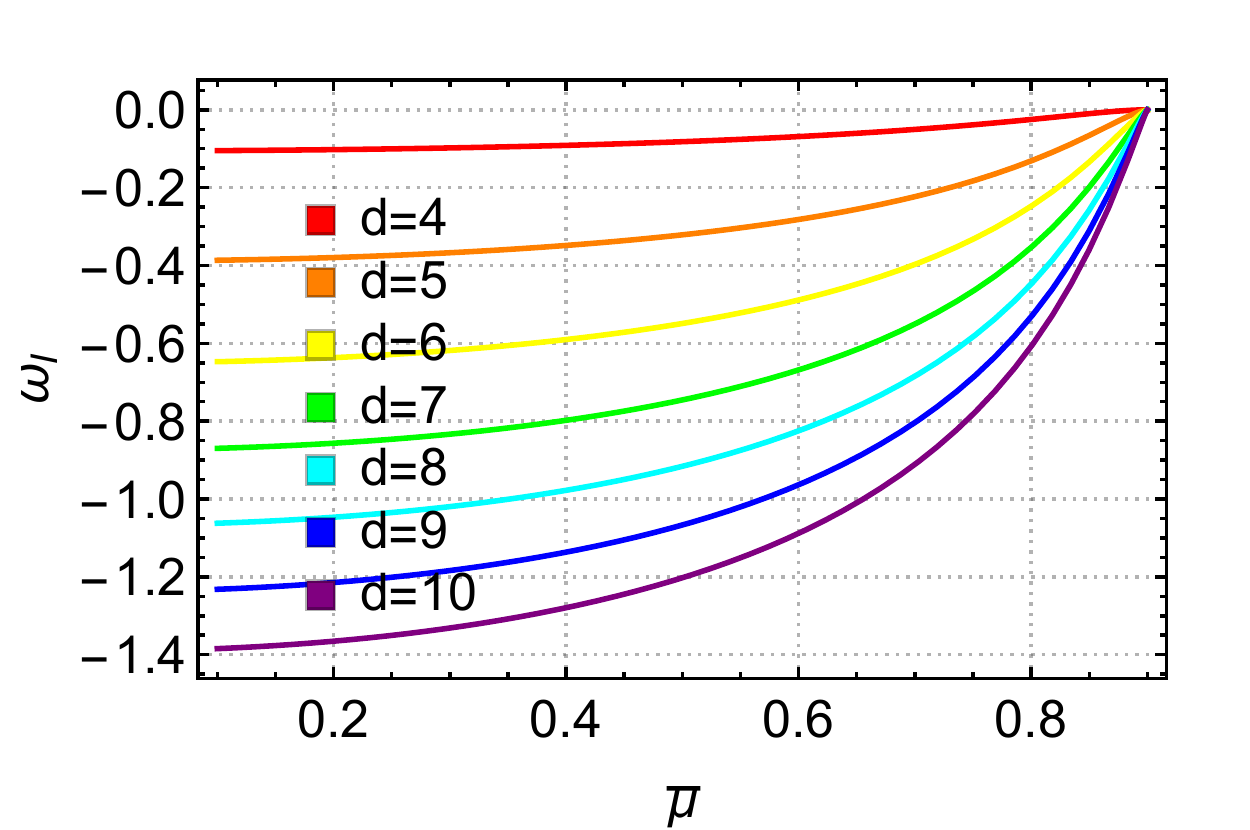}
   \caption{The image part of the quasinormal resonance frequency for $\bar{Q}=0.9, q=0.1, M=1, n=0$.}
   \label{lambdadiagram}
\end{figure}

Based on the quasinormal resonance frequencies (\ref{ome}) we have obtained, we can determine the characteristic timescale $\tau_{relax}$ associated with the linear relaxation dynamics of the charged scalar field in the background of the higher-dimensional RN spacetime, as 
\begin{equation}
\tau_{relax}=\frac{1}{\omega_{I0}},
\end{equation}
where $\omega_{I0}$ is the image part of the fundamental quasinormal frequency ($n=0$). As shown in the Fig. \ref{lambdadiagram}, the image parts of the quasinormal frequencies approach to zero when $\bar{\mu}/\bar{Q}\rightarrow 1^{-}$ for all spacetime dimensions. As a result, we can conclude that
\begin{equation}
\tau_{relax}\rightarrow \infty,~~\text{for}~~\frac{2(d-3)\mu M}{(d-2)qQ}\rightarrow 1^{-},
\end{equation}
which implies nothing else but the charged massive scalar fields can exist outside the higher-dimensional RN black holes for a rather long relaxation time, though exponentially decay and finally elapse.

As shown above in (\ref{ome}), we have successfully derived the real and image parts of the quasinormal resonance frequency for the charged massive scalar field in the background of higher-dimensional RN spacetime. However, one should note that our analytical results are based on not only the conditions (\ref{eik}), (\ref{rbiggeri}), but also higher order corrected terms in (\ref{wkb}) having been neglected. The higher order terms including up to fourth-order derivatives are \cite{Berti:2009kk,Hod:2016jqt}
\begin{equation}\begin{aligned}
\Lambda (n)=\frac{1}{\sqrt{2V_{0}^{(2)}}} &\left[\frac{1+(2n+1)^2}{32}\cdot \frac{V_{0}^{(4)}}{V_{0}^{(2)}}\right.\\&\left.-\frac{28+60(2n+1)^2}{1152}\cdot \left(\frac{V_{0}^{(3)}}{V_{0}^{(2)}}\right)^2\right].
\end{aligned}\end{equation}
The WKB approximation (\ref{wkb}) is valid on the condition that 
\begin{equation}
\frac{\Lambda (n)}{n+\frac{1}{2}}\ll 1.
\end{equation}
By tedious calculation, we find that the condition
\begin{equation}
q Q \sqrt{2 (\text{d}-3)-(\text{d}-2) \bar{\mu}^2}\gg n+\frac{1}{2},
\end{equation}
should be fulfilled to make our WKB approximation (\ref{wkb}) valid.

\section*{Acknowledgements}
This work is supported by the National Natural Science Foundation of China (NSFC) with Grants No. 11235003, 11775022, 11375026, 11475179 and 11675015. We would like to thank Hongbao Zhang for his valuable discussions.

\appendix

\section{The eikonal large-mass regime}\label{appa}
The mass of the $d$-dimensional RN black hole can be expressed as
\begin{equation}
r_+ =\left(\frac{4 \pi}{\epsilon} \right)^{\frac{1}{d-3}}\left(\frac{2 M}{d-2}+\sqrt{\frac{4 M^2}{(d-2)^2}-\frac{2 Q^2}{d^2-5 d+6}}\right)^{\frac{1}{d-3}},
\end{equation}
then we can know
\begin{equation}
r_+ \geqslant \left[\frac{8 \pi  M}{(d-2) \epsilon }\right]^{\frac{1}{d-3}},~
\end{equation}
and
\begin{equation}
M^2 \geqslant \frac{(d-2) Q^2}{2 (d-3)}.
\end{equation}
So we have
\begin{equation}
\frac{K_l}{\mu^2 r^2}<\frac{l(d+l-3)}{\left(\mu  \left(\frac{8 \pi  M}{(d-2) \epsilon }\right)^{\frac{1}{d-3}}\right)^2}\sim 0^+,
\end{equation}
\begin{equation}
\frac{d^2-6 d+8}{4\mu^2 r^2}<\frac{d^2 -6d+8}{\left(\mu  \left(\frac{8 \pi  M}{(d-2) \epsilon }\right)^{\frac{1}{d-3}}\right)^2}\sim 0^+,
\end{equation}
\begin{equation}
\frac{4 \pi  (d-2) M}{\mu ^2 \epsilon  r^{d-1}}<\frac{(d-2)^2}{2\mu^2 r^2}<\frac{(d-2)^2}{2}\frac{1}{\left(\mu  \left(\frac{8 \pi  M}{(d-2) \epsilon }\right)^{\frac{1}{d-3}}\right)^2}\sim 0^+,
\end{equation}
\begin{equation}
\frac{8 \pi ^2 (3 d-8) Q^2}{\mu^2 (d-3) \epsilon ^2 r^{2 d-4}}<\frac{(3d-8)(d-2)}{4\mu^2 r^2}<\frac{(3d-8)(d-2)}{4\left(\mu  \left(\frac{8 \pi  M}{(d-2) \epsilon }\right)^{\frac{1}{d-3}}\right)^2}\sim 0^+,
\end{equation}
in the eikonal large-mass regime
\begin{equation}
  \mu\left(\frac{8\pi}{(d-2)\epsilon }M\right)^{\frac{1}{d-3}}\gg d+l-3.
\end{equation}

\section{Several expressions of the image parts of the quasinormal frequencies }\label{appb}
As instances, we here show some detailed expressions of the image part of the quasinormal frequencies. For $d=4$,
\begin{equation}
\omega_I=-\frac{(2 n+1) \left(\bar{Q}^2-1\right) \sqrt{\frac{\left(1-\bar{\mu }^2+\sqrt{\left(1-\bar{\mu }^2\right) \left(1-\bar{Q}^2\right)}\right)^4}{\left(1-\bar{\mu }^2\right)^3}}}{2 M \bar{Q}^4 \sqrt{\left(1-\bar{\mu }^2\right) \left(1-\bar{Q}^2\right)}}.
\end{equation}
For $d=5$,
\begin{equation}\begin{aligned}
\omega_I =&-\frac{\sqrt{3 \pi } \left(n+\frac{1}{2}\right) \sqrt{1-\bar{Q}^2}}{\sqrt{M} \bar{Q}^3}\\&\times \sqrt{\frac{\left(\bar{Q}^2-\bar{\mu }^2\right)^3}{\left(1-\bar{\mu }^2\right) \left(4-\bar{\mu }^2-3 \bar{Q}^2\right)\mathcal{C}_5 }},
\end{aligned}\end{equation}
where 
\begin{equation}
\mathcal{C}_5 =\left(1-\bar{\mu }^2\right)+\sqrt{\left(1-\bar{\mu }^2\right) \left(1-\bar{Q}^2\right)}.
\end{equation}
For $d=6$,
\begin{equation}
\omega_I =-\frac{3^{2/3} \sqrt[3]{\frac{\pi }{2}} (2 n+1) \left(1-\bar{Q}^2\right)}{\bar{Q}^2 \sqrt{\left(1-\bar{\mu }^2\right) \left(1-\bar{Q}^2\right)}}\times \sqrt{\mathcal{C}_{61}\mathcal{C}_{62}},
\end{equation}
where 
\begin{equation}
\mathcal{C}_{61} =\left(\frac{\bar{Q}^2-\bar{\mu }^2}{M \bar{Q}^2 \left(-\bar{\mu }^2+\sqrt{\left(1-\bar{\mu }^2\right) \left(1-\bar{Q}^2\right)}+1\right)}\right)^{2/3},
\end{equation}
\begin{equation}
\mathcal{C}_{62}=\frac{\left(\bar{Q}^2-\bar{\mu }^2\right)^2}{-\bar{\mu }^2+2 \sqrt{\left(1-\bar{\mu }^2\right) \left(1-\bar{Q}^2\right)}-\bar{Q}^2+2}. 
\end{equation}
For d=7,
\begin{equation}
\omega_I =-\frac{\sqrt[4]{10} \sqrt{\pi } (2 n+1) \left(1-\bar{Q}^2\right)}{\bar{Q}^{5/2} \sqrt{\left(1-\bar{\mu }^2\right) \left(1-\bar{Q}^2\right)}}\times \sqrt{\mathcal{C}_{71}\mathcal{C}_{72}},
\end{equation}
where 
\begin{equation}
\mathcal{C}_{71} =\sqrt{\frac{\bar{Q}^2-\bar{\mu }^2}{M \left(-\bar{\mu }^2+\sqrt{\left(1-\bar{\mu }^2\right) \left(1-\bar{Q}^2\right)}+1\right)}},
\end{equation}
\begin{equation}
\mathcal{C}_{72}=\frac{\left(\bar{Q}^2-\bar{\mu }^2\right)^2}{-\bar{\mu }^2+2 \sqrt{\left(1-\bar{\mu }^2\right) \left(1-\bar{Q}^2\right)}-\bar{Q}^2+2}.
\end{equation}

\section*{References}
\bibliographystyle{elsarticle-num}
\bibliography{Notes}

\end{document}